\documentclass[12pt]{article}
\usepackage{amsmath, amssymb}
\usepackage{graphicx}
\usepackage{hyperref}
\usepackage{dirtytalk}
\usepackage[a4paper, total={7in, 10in}]{geometry}

\title{\textbf{A critique of one of Galileo's mental experiments and an explanation on why bodies fall with equal acceleration through the vacuum}}
\author{Leonardo Levinas \\ Universidad de Buenos Aires-CONICET Argentina \\ \href{leolevinas@gmail.com}{leolevinas@gmail.com} }
\date{}
\begin{document}

\maketitle

\begin{abstract}
In the famous thought experiment studied in this article, Galileo attempted to refute the Aristotelian hypothesis that heavier bodies should fall more quickly than lighter ones. After pointing out some inconsistencies in Galileo's approach, we show, through the design of two alternative but equivalent experiments, that from his imaginary experiment, it is not possible to reach the conclusion that all bodies fall simultaneously. We show why, to explain the result of this type of experience, it is necessary to establish the equivalence between inertial and gravitational masses derived exclusively from experience.

Pacs numbers:  01.55.+b \href{https://ufn.ru/en/pacs/01.55.+b/}{General physics}, 01.65.+g \href{https://ufn.ru/en/pacs/01.65.+g/}{History of science}, \href{https://ufn.ru/en/pacs/01.70.+w/}{Philosophy of science} 01.70.+w, 04.20.-q \href{https://ufn.ru/en/pacs/04.20.-q/}{Classical general relativity}.


\end{abstract}
\section{Introduction}
The problem of freely-falling bodies and the thought experiment that we will examine here was exposed by Galileo in his \textit{Discorsi} of 1638. Galileo's intention was to refute Aristotle's assertion that heavier bodies should fall faster than lighter ones. This experiment has been thoroughly analyzed in the literature; among other works, we highlight those of Crombie (1957), Casper (1977), Koyré (1968, 1978), Brown (1991), Norton (1996), Gendler (1998), Palmieri (2005), and more recently, those of Grundmann (2018), El Skaf (2018) and Mondragón (2020). However, our approach to the problem differs from those of all those authors.

Our work is organized as follows. First, in \ref{section2}. we intend to show why Galileo's reasoning is inconsistent and invalid to refute the Aristotelian theory about falling bodies with different weights. In \ref{section3}  we strengthen our argument based on the design of two alternative experiments to those of Galileo. In \ref{section4} we will determine why the result arises that all bodies moving in a vacuum within a uniform gravitational field must fall with the same acceleration. Finally, in section \ref{section5} we offer a summary of the most important conclusions we have reached.

\section{A critical review of galileo's mental experiment 
 }\label{section2}

\subsection{\textbf{Galileo's Mental Experiment}
}\label{section2.1}
In the \textit{Discorsi,} (Galileo, 1638, 63) Salviati, the character who defends the ideas of Galileo, chooses an Aristotelian hypothesis that he considers erroneous. That hypothesis is: if one body weighs more than another, the speed at which it falls towards the ground will be greater. The purpose of the experiment Galileo imagined is to show that the Aristotelian hypothesis leads to two contradictory conclusions, in the ultimate aim of demonstrating that bodies actually fall simultaneously. The experiment consists of the following: 

Consider the Aristotelian hypothesis on the free fall of bodies: if we have a body A (B) whose weight is \textit{W\textsubscript{A}} (\textit{W\textsubscript{B}}) being that \textit{W\textsubscript{A} $>$ W\textsubscript{B}} and we let them fall simultaneously from the same height, then A will fall with more velocity than B; that is, it will take less time to reach the ground. In Galilean terms, the Aristotelian hypothesis indicates that A would fall faster than B: \textit{v\textsubscript{A}} \textit{$>$ v\textsubscript{B}}. 

Following Salviati (defender of Galilean ideas), suppose both bodies are united by means of a rigid mechanism; then, from the hypothesis in question \textit{two} conclusions could be drawn: 

(a) The bodies A and B should fall together at an intermediate velocity, between velocities \textit{v\textsubscript{A}} and \textit{v\textsubscript{B.}} Its velocity would in-between because body B would tend to brake body A, while body A would tend to accelerate body B. Accordingly: \textit{v\textsubscript{A} $>$ v\textsubscript{A+B }$>$ v\textsubscript{B}}. 

(b) On the other hand, the system composed of A and B should weigh \textit{W\textsubscript{A+B}} \textit{= W\textsubscript{A}+ W\textsubscript{B}}. Therefore, this new body should fall faster than the heavier body A:  \textit{v\textsubscript{A+B} $>$ v\textsubscript{A} }. 

Then, according to Salviati, the only way to bridge the contradiction between (a) and (b) would be to reject the Aristotelian hypothesis and admit that all bodies must fall from the same height in the same amount of time. That is: the time of each drop must be: \textit{t\textsubscript{A+B}= t\textsubscript{A}= t\textsubscript{B}}.

While we know experience shows that the bodies would fall with the same acceleration in all cases, said experience is not derived from a contradiction between the two conclusions (a) and (b) from the mental experiment of the \textit{Discorsi}. In fact, in the same text, although based on \textit{real} experiences this time, Galileo will express the idea that all bodies without resistance would fall at the same speed, this time based on experimental data. He will argue that, since the final velocity in air of two bodies composed of different materials tends to be the same, within the constraint of a non-resistant medium, i.e., in vacuum, the velocity of all bodies should be the same (Galileo, 1638, 71-2).

\subsection{\textbf{Some inconsistencies in Galileo's reasoning}} \label{section2.2}

Despite Galileo's undoubted ingenuity, his mental experiment does not really lead to a refutation of Aristotelian ideas. In fact, when Aristotle held that a heavier body would fall in less time than a lighter one he meant \textit{natural} movements, that is, movements in which heavy bodies would seek their \textit{natural} place in the center of the Earth. Therefore, an important consideration that we can make in defense of the Aristotelian point of view, is to recognize that when we join two bodies of different weights, if one accelerates the other and the other tends to slow down the first (situation (a)), the movement would no longer be\textit{ natural} for either body. This distinction alone invalidates Galileo's argument—since, when in the first situation the two bodies are linked, the conditions offered for a free fall would not be met for either body. In other words, neither of the two bodies would follow a natural movement but a forced one: the lighter body would exert a force of resistance on the heavier one, and, conversely, the heavier would force the lighter one to fall with greater speed. 

On the other hand, Galileo assumes that in the second situation (b), bodies A and B form a single body C whose weight would be the sum of the weights of each. That is, it would have the weight of A plus the weight of B, so that the ``part" A of C and the ``part" B of C would fall with the same speed, as in the previous case, but this time without ``interfering" with each other since they would be part of the same body (rigid). This situation is completely different from that of (a). The latter contradicts something indicated by Galileo himself in his \textit{Discorsi }(Galileo, 1638, 63-4), where he firmly maintains that in this last situation, A and B would fall simultaneously, but without adding their weights. In other words, Galileo recognized that two bodies behaved differently when they fell together than when they were weighed together. According to Galileo, a smaller stone adds weight to the larger one when they are together at rest (for example on a scale), which does not happen when they fall. Thus, B does not brake A, just as Galileo himself suggests. 

Later we will see that this is so because when the bodies fall freely they do \textit{not} have weight. 

\section{Two alternative experiments showing why Galileo's mental experiment does \textit{not }refute the aristotelian hypothesis
} \label{section3}

\subsection{The case of two falling spheres 
}
\begin{figure}[h]
    \centering
    \includegraphics[width=0.2\linewidth]{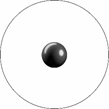}
    \caption{A small sphere B in the center of a heavier hollow sphere A. }
    \label{fig1}
\end{figure}

In our first mental experiment, we slightly modified the mental experiment suggested by Galileo in \textit{Discorsi}. We have a hollow, heavy iron sphere and a smaller, lighter wooden sphere. Suppose the initial state presents the situation shown in Fig. \ref{fig1}

Now let's let both spheres fall. In accordance with the Aristotelian hypothesis – and without the need to employ any formalism – we should agree that at some point the situation shown in Fig. \ref{fig2} will take place: 

\begin{figure}[h]
    \centering
    \includegraphics[width=0.2\linewidth]{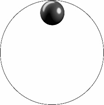}
    \caption{Sphere B in contact with the hollow, heavier sphere A.}
    \label{fig2}
\end{figure}
We would still have two bodies: body A made of iron and body B of wood. If Aristotle's hypothesis is correct, once A and B are in contact, they will begin falling together. B would tend to decelerate A while A would tend to accelerate B and the velocity of the whole would be between that of A and B falling separately. That is to say, (a) would be fulfilled (see \ref{section2.1}). However, for both A and B, the conditions indicated by Aristotle would \textit{not} be met since the rate of acceleration of each body, attributed exclusively to its respective weight, would be affected by the interaction with the other body. That is, the movements of A and B would no longer be natural but forced. However, if we weighed A and B separately, or even together, on a scale we would get a weight of \textit{W\textsubscript{A}+W\textsubscript{B}}. 

Now, if we start from a \textit{different} situation in which both bodies were already united from the beginning forming, in fact, a single body whose weight was the weight of A plus the weight of B, then, according to the Aristotelian hypothesis, the result should be that this “new" body would fall faster than body A falling alone. But in this case, it would be the \textit{natural }fall of a (single) body C and (b) would be fulfilled, which would not contradict the Aristotelian hypothesis.

\subsection{The case of two charges falling on an electric field}

To show in another way why Galileo's reasoning is invalid – in particular, why conclusion (b) is not derived from the Aristotelian hypothesis –, we present another mental experiment that also involves the fall of bodies at rest from a certain height, equivalent to the experiment suggested by Salviati. The only difference is that, in modern terminology, the fall of bodies occurs not only in the presence of a gravitational field but also in the presence of an electric field.

Consider the following, always in the presence of a vacuum: 

There are two bodies, A and B, with the same weight \textit{W\textsubscript{A} = W\textsubscript{B}}, with positive charges q\textsubscript{A} and q\textsubscript{B, }respectively, (q\textsubscript{A} $>$ q\textsubscript{B} $>$ 0). We drop them from the same height \textit{r} with respect to the surface of a sphere with the size of the Earth, mass \textit{M} and a large negative charge \textit{Q}. Our question is about how they will fall.

Let's try the following “Aristotelian" hypothesis: if we have two bodies A and B, with the same weight but with different positive charges q\textsubscript{A} and q\textsubscript{B} (q\textsubscript{A }$>$ q\textsubscript{B }$>$ 0), under the influence of a uniform electric field produced by a large and close negative charge \textit{Q}, then A will fall with a greater rate of acceleration than B. 

Now suppose we tie both bodies together using an insulating, rigid ligature.\footnote{The rigid ligature would prevent the two charged bodies from repelling each other and would cause the ensemble to behave like a rigid body.}Then, following Galileo's reasoning, we should consider the following two possible conclusions: 

(a’) Body A will tend to accelerate body B, while B will tend to slow down body A. The result should be an intermediate system-wide acceleration: \textit{a\textsubscript{A} $>$ a\textsubscript{A+B} $>$ a\textsubscript{B} }. 

(b’) This new system has a higher charge than body A; consequently, following the hypothesis: \textit{a\textsubscript{$A+B$} $>$ a\textsubscript{B} }.

Again, the only way to resolve the contradiction between the two conclusions would be that \textit{a\textsubscript{A+B }= a\textsubscript{A }= a\textsubscript{B}} . In other words, we should conclude that all charged bodies (with identical weight) must move with the same acceleration in the presence of an electric field. 

Now, considering what experience shows for the case of charged bodies, we would see that the Aristotelian hypothesis applied to these bodies would be true. But it would also show us that (a’) is fulfilled and (b’) is not! The reason is simple: (b’) is not derived from the hypothesis.

To understand why, let us study the explanation of this phenomenon using the law of Coulomb, the law of gravitational attraction, and Newton's second law (unknown to Galileo and, of course, Salviati). Using these laws, we have (charges are in Gaussian units): 

\begin{equation}\label{eq1}
a_A = \frac{q_A Q}{m \left( r + R \right)^2} + \frac{GM}{m \left( r + R \right)^2}
\end{equation}
\begin{equation} \label{eq2}
    a_B = \frac{q_B Q}{m \left( r + R \right)^2} + \frac{GM}{m \left( r + R \right)^2}
\end{equation}

 where \textit{R} is the radius of a charged “Earth” (\textit{r$\ll$R}) and \textit{G} is the gravitational constant. From Eqs. \ref{eq1} and \ref{eq2} we see that in line with the hypothesis: a\textsubscript{A }$>$ a\textsubscript{B }. \textit{This result comes from experimentally confirmed laws}. Let's find the acceleration to A+B of the compound system A+B, where the charged body A is above the charged body B, both bound by a rigid non-conductive cable of negligible weight. With the appropriate initial conditions, we would have both bodies A and B moving from rest and towards the “Earth" at the same rate of acceleration as A+B since they would fall simultaneously: 
\begin{equation}\label{eq3}
    ma_{A+B} = \frac{q[A]Q}{(r + R)^2} + \frac{GmM}{(r + R)^2} - T,
\end{equation}

\begin{equation}\label{eq4}
    ma_{A+B} = \frac{q[B]Q}{(r + R)^2} + \frac{GmM}{(r + R)^2} + T,
\end{equation}

where \textit{T} is the ligation force (\textit{T} “absorbs" the electric repulsion force between both positive charges q\textsubscript{ A} and q\textsubscript{ B}).

Solving the system of equations \ref{eq3}$-$\ref{eq4}, we find that: 
\begin{equation}
    a_{A+B} = \frac{(q_A + q_B)Q}{2m \, (r + R)^2} + \frac{GM}{(r + R)^2}
\end{equation}
Taking into account that $r\ll R$, $GM/R=g$  where $g$ is the value of the acceleration of gravity on the surface of the “Earth", and that $\frac{Q}{(r+R)^2} =E$, where $E$ is the value of the electric field created by the charge $Q$ on the surface of the “Earth", we obtain:

\begin{equation}\label{eq6}
    a_{A+B} = \frac{(q_A + q_B)E}{2m} + g
\end{equation}
As we can see, this is equivalent to making use of the eq. \ref{eq1} for a body with charge \textit{q\textsubscript{A}+q\textsubscript{B}} and a mass of 2\textit{m}. Since 2\textit{q\textsubscript{A }$>$ q\textsubscript{A}+q\textsubscript{B} $>$} 2\textit{q\textsubscript{B}}, we observe that body B effectively slows body A while body A accelerates body B. \footnote{The acceleration \ref{eq6} would also be valid in the event that the cable was conductive and there was a new distribution of the charges such that the total load of the falling body was \textit{q\textsubscript{(tot)} = q\textsubscript{A}+q\textsubscript{B}} .}

In conclusion: (a’) is true and (b’) is false. Conclusion (b’) would be valid if system A+B acted as a body of mass \textit{m}, and not of mass 2\textit{m}. But system A+B not only has more charge than A, it also has more mass, and the ratio between both amounts is less than that of A. In this combined body, as in any body, the acceleration due to electrical attraction is proportional to its total charge, while the resistance to being accelerated (to change its state of motion) is proportional to the total inertial mass.\footnote{Note that if charges q\textsubscript{A} and q\textsubscript{B} were equal but opposite, the electric field created by Q would have no effect on the system and its acceleration would be equal to \textit{g}.} A hypothesis similar to that of Aristotle would be fulfilled where simply the terms “weight", “heavy", “heavier", etc., were replaced by the terms “charge", “charged", “more charged", etc. The hypothesis would be: “If two bodies that only differ in their charges are dropped, then the one that has the greater charge will fall with greater acceleration"; while only the first of the conclusions suggested by Galileo in his mental experiment would be fulfilled, this is conclusion (a’): “The combined bodies would fall with an intermediate acceleration to those corresponding to the two bodies falling separately", and conclusion (b’) would \textit{not} be fulfilled.

\section{ Why do all the bodies, when free falling in a void, do so with the same rate of acceleration? 
} \label{section4}

\subsection{The Word of Experience 
}
All charged bodies would fall in the same way \textit{regardless} of their charge if for each body: 
\begin{align}
q &= \alpha m_I  \label{6a}\\
m_I &= \beta m_G = m  \label{6b}
\end{align}
where $\alpha$ and $\beta$ would be two constants ($\alpha$ with units = $[q]/[m]$ ) that do not depend on the field. The magnitudes \textit{m\textsubscript{I}} and \textit{m\textsubscript{G}} are the \textit{inertial }and \textit{gravitational masses,} respectively (in our equations we have taken $\beta$=1). However, we know through experimentation that this is not the case in nature: although \ref{6b} is true, \ref{6a} it is \textit{not}. Unlike gravitational mass, charge is not proportional to inertial mass: that is, charge and mass are properties of mutually independent bodies.\footnote{It is important to notice that there is \textit{no} “charge particle", that is, the amount of charge in a body is independent of its mass and volume, as a result of which the total charge of a body with a given volume is not divisible into “charge particles", as it is in mass particles, from which we would obtain the total mass of the body.} So much so that when we join two charged bodies, we combine their inertial and gravitational masses in the same proportion:

\begin{equation}
m_{AI} + \frac{m_{BI}}{m_{A G}} + m_{B G} = \frac{m_{AI}}{m_{A G}} = \frac{m_{BI}}{m_{B G}}
\end{equation}
but we do \textit{not} combine their inertial masses and their charge in identical proportion. For the case of our example \ref{section2.2}: 

\begin{equation}
    (q_A>q_B , m_A=m_B):   \frac{q_A}{m_A} > \frac{q_A+q_B}{m_A+m_B} > \frac{q_B}{m_B}.
\end{equation}

The acceleration of the free-falling bodies in a vacuum \textit{does not} depend on inertial mass since weight is proportional to it. Additionally, Coulomb's law tells us that the electric force \textit{does not depend} on mass either; therefore, the acceleration of a charged body in the presence of an electric field \textit{does} depend on it. \footnote{All bodies, charged or not, would fall in the same way if it were always true that: \textit{q\textsubscript{1}.\textsubscript{.}q\textsubscript{2}/r $\ll$ km\textsubscript{1}.m\textsubscript{2}/r}, but, again, this is not the case in nature: the gravitational constant is so small that, for example, two electrons at rest are repelled (electrically) with a force that is 4x10\textsuperscript{42} times the gravitational force that attracts them.}

\subsection{Why do bodies in the vacuum fall with equal rates of acceleration?}

Both Aristotle and Galileo recognized the fact that on the surface of the Earth it was more difficult to move a larger body than a smaller one of the same composition, either to raise it or to transport it. Today we accept that the variable that intervenes when a force is applied to a body is the \textit{inertial mass}, which opposes a change in its state of movement.

Bodies fall with equal acceleration in a vacuum because, on the one hand, i) the force of gravitational pull exerted by the Earth on a heavier body is greater than that exerted by the Earth on a lighter one, since the gravitational mass of the former is greater than that of the latter; but, in turn, ii) this occurs in the same proportion in which the Earth has more “difficulty" to move the heavier body than the lighter one since the former has more inertial mass – which translates into a greater resistance against changing its state of movement – than the latter. This is equivalent to assuming that the inertial and gravitational masses are equal (or rather, proportional).

Precisely, experience shows the equivalence between inertial and gravitational masses. This is manifested in the equation \textit{force = mass×acceleration}; in our case: \textit{m\textsubscript{G}×g = m\textsubscript{I}×a}. If we take \textit{m\textsubscript{I} = m\textsubscript{G} = m }, then \textit{a = g} for all bodies in free fall in a vacuum. On the one hand, a force proportional to the magnitude \textit{m\textsubscript{G }}is exerted, and, on the other, the body offers a resistance proportional to \textit{m\textsubscript{l}}, as if the body had two masses that were “activated" simultaneously. The important point is that the evidence that inertial mass \textit{m\textsubscript{I }}and the gravitational mass \textit{m\textsubscript{G}} are proportional was provided by \textit{real} experiments such as those designed by Galileo using pendulums or inclined planes, or the more sophisticated ones by Eötvos (1922) or Dicke (1967) which form the basis for establishing the Weak Equivalence Principle (WEP), one of the foundations of Einstein's theory of General Relativity. The WEP states that the universe line of a free-falling (freely gravitating) test particle is independent of its composition and structure (Misner et al., 1973).

It should be noted that in real conditions, that is, in a dense medium such as air, the relationship between the weight, the air resistance and the acceleration of the falling body, is represented, first, in terms of velocity, and then in terms of Newton's second law, by the following differential equation: \textit{F = m dv/dt = mg - kv}, where \textit{k }is the resistance coefficient, which leads to the following solution for the falling velocity \textit{v }[with initial velocity $v(t=0)=0$]: $v(t)=(mg/k)-(mg/k)e^{(-k/m)t}$ ]. From the solution it then follows that when time \textit{t} is large enough, the body reaches a constant final velocity. This terminal velocity is \textit{v\textsubscript{term} = mg/k}. The final velocity, then, is proportional to the weight \textit{mg}, and the heavier bodies, in a medium such as air, fall with a higher final velocity. This means that under real conditions and for fall times that are long enough, Aristotle was right.\footnote{Aristotle held that the rate of fall of a heavy body was proportional to its weight and inversely proportional to the resistance of the medium, and invoked this fact as one of the reasons why a vacuum could not exist: if the resistance of the medium were zero (vacuum), then the falling velocities should be infinite (Aristotle, Book IV, 8, 62)}

The best way to interpret why the behavior of bodies in free fall in a vacuum does not depend on their weight is to think that when a body falls, it does not have weight. Clearly, \textit{m\textsubscript{G}} acts when the body “rests”, for example on a horizontal plane. Then, the Earth exerts a force on it which is the weight \textit{P = m\textsubscript{G}.g}, where \textit{g} is the acceleration of gravity, and it is balanced by reaction, which is none other than the “normal force" \textit{N = –m\textsubscript{G}.g;} here the phenomenon is static. When we move a body horizontally (in a horizontal plane without friction), the force needed to accelerate it with an acceleration \textit{a} is \textit{m\textsubscript{I}.a}, and the phenomenon becomes dynamic. As for the case in the middle, an inclined plane without friction by which the body would fall, the force applied in the direction of movement is \textit{m\textsubscript{I}.g.}$\sin \alpha$ (where $\alpha$ is the inclination angle of the plane), while the force perpendicular to the movement is \textit{m\textsubscript{I}.g.}$\cos \alpha$ in the direction in which there is no movement, that is: the latter force does not intervene dynamically and represents the weight of the body on the inclined plane. 

Suppose, now, that the inclined plane is, in reality, an inclined weighing scale. If $\alpha = 0$  then the plane would be horizontal and the scale would register a weight \textit{m\textsubscript{l}.g} greater the greater \textit{m\textsubscript{I}} was. As a result, there would be no (horizontal) movement. However, if   $\alpha = 90$°, then the plane would be vertical and there would be a vertical free fall movement, but the scale would not register any weight. 

Bodies fall with acceleration due to the gravitational pull of the Earth and do so with the same acceleration because in a scale that were connected to these bodies (that is, also in free fall), nothing would be recorded: \textit{bodies in free fall do not have weight}. 

At this point it is important to note that for Newton himself, inertia was activated only when a force was applied to the body: 

\begin{quote}
 Because of the inertia of matter, every body is only with difficulty put out of its state either of resting or of moving. Consequently, inherent force [\textit{vis insita}] may also be called by the very significant name of force of inertia [\textit{vis inertiae }]. Moreover, a body exerts this force only during a change of its state, caused by another force impressed upon it (…) (Newton, 1687, 404).
 \end{quote}

Something similar occurs with the force of weight in the sense that weight does \textit{not} manifest itself when the body falls and is activated when it has a resistance, for example, on a scale. That analogous behavior is why WEP indicates that the movement of bodies in free fall, in a freely gravitational system, is equivalent to their motion in an inertial system; no scale would prevent them from falling since the scale itself would fall with the same acceleration. In other words, while something falls, it is not in an act of “weighing", just as something to which no net force was applied would not be in an act of resisting a force to change its state of movement. 

\section{Conclusions}\label{section5}

We can piece the conclusions together in the following points: 

1) The \textit{Discorsi }mental experiment does not refute the Aristotelian hypothesis that heavier bodies fall faster than lighter ones. In our criticism of Galileo's mental experiment we show, first of all, that in its interpretation and in the use of the Aristotelian hypothesis, Galileo incurs two “omissions". As for the first conclusion that we have called (a), “The system composed of a body A heavier than body B would fall with an intermediate acceleration with compared to those of A and B respectively," it contradicts the Aristotelian idea of natural fall since if the bodies interact, according to Aristotle, their fall would be forced and therefore unnatural. As for the second conclusion that we have called (b), “The composite system would be a heavier body than A so it should fall faster than it," it would contradict Galileo's own assertion that bodies A and B in the situation suggested in (a) would not fall independently or with the same acceleration, that is, without affecting each other, since they would make up a single body that on a scale would weigh the sum of the weights of A and B. In the analysis of his mental experiment, we have seen that Galileo “omits" the fact that he himself discovered—that the bodies in the void would fall with the same acceleration, such that a body B linked to a heavier body A would not influence its fall since both falls would be independent, which contradicts (a). The A + B system weighs the sum of the value of the weights A and B when weighed on a scale. When the body falls, it does not weigh, something that in the \textit{Discorsi }Galileo himself admits. 

2) In an experiment with a design equivalent to that of \textit{Discorsi, }which consists in imagining what would happen in the case of a small compact sphere initially located in the center of another hollow sphere, it is shown that whenever the Aristotelian hypothesis was correct, we would obtain (i) that the inner sphere should contact the hollow sphere, and (ii) that the lighter one (the smaller one) would slow down the large and hollow sphere (always according to the Aristotelian hypothesis); and conversely, that the larger sphere would tend to accelerate the smaller one, so that the whole would fall with an intermediate acceleration with respect to those that both spheres would carry before contact. 

3) To show in another way why Galileo's reasoning is invalid – in particular, why conclusion (b) is not derived from the Aristotelian hypothesis –, we have presented another mental experiment equivalent to the experiment suggested by Salviati. Using empirical laws such as Coulomb's for two linked bodies with positive charges, it is observed that in a free fall towards a “Earth" with a large charge and of the opposite sign, only the first of the conclusions (a’) would be extracted from an Aristotelian hypothesis, that is: the bodies with a greater (positive) charge, would fall faster towards an “Earth" charged (negatively), and the “compound" body would fall with an intermediate speed.

4) The reason the bodies fall simultaneously is the equivalence between inertial mass and gravitational mass which leads to the following: while the force exerted by the Earth on a heavier body (that is, which has a greater gravitational mass) is greater than on a lighter body, the resistance (due to the inertial mass) offered by a heavier body is greater in the same proportion as it is for a lighter body. This equivalence of the masses can only be corroborated experimentally and forms the basis of the so-called Weak Equivalence Principle. 

5) In real conditions, like in a dense medium (like air), a body heavier than another of the same size and shape would take less time to fall than the lighter one (as pointed out by Aristotle who, by the way, did not believe in the existence of the void). Both would reach a final constant speed, but the speed of the heavier one would be greater than that of the lighter one. In turn, the \textbf{real} experiments carried out by Galileo under \textbf{real} conditions showed that going to the limit of a medium without resistance (the vacuum) all bodies would tend to fall with identical acceleration, regardless of their weight.

\subsection*{\textbf{Acknowledgments}}

To Nathalie Deruelle, Rafael Ferraro, Alejandro Cassini and Jorge Miraglia for their respective critical readings of the manuscript and for their invaluable contributions for a better exposition of the ideas and results. This work was possible thanks to funding from the UBACyT 20020170100124BA Project of the University of Buenos Aires, and the PIP 586 Project of the National Council for Scientific and Technical Research (CONICET).

\end{document}